\documentclass[twocolumn,preprintnumbers,amsmath,amssymb,superscriptaddress,prb]{revtex4}
\usepackage{graphicx}
\usepackage{longtable}
\usepackage{verbatim}

\usepackage{color}

\newcommand{\notekj}[1]{}

\begin{document}
	\title{Long-period suspended silicon Bragg grating filter for hybrid near- and mid-infrared operation}
	\author{Carlos Alonso-Ramos$^{1}$, Xavier Le Roux$^{1}$, Daniel Benedikovic$^{1}$, Vladyslav Vakarin$^{1}$, Elena Dur\'{a}n-Valdeiglesias$^{1}$, Diego P\'{e}rez-Galacho$^{1}$, Eric Cassan$^{1}$, Delphine Marris-Morini$^{1}$, Pavel Cheben$^{2}$, Laurent Vivien}
	\address{
		Centre for Nanoscience and Nanotechnology, CNRS, Univ. Paris-Sud, Universite Paris-Saclay, C2N - Orsay, 91405 Orsay cedex, France
		\\
		$^2$National Research Council, Ottawa, K1A 0R6, Canada
	}

	\date{\today}
	

	\begin{abstract} 
		The large transparency window of silicon, covering the $1.1-8\,\mu\mathrm{m}$ wavelength range, makes it a promising platform for the implementation of photothermal-based absorption spectrometers. These devices indirectly sense absorption in the mid-infrared (MIR) by using near-infrared (NIR) wavelengths, thereby enabling the realization of MIR absorption spectrometers without the need for MIR photodetectors. Nevertheless, due to their comparatively large index contrast and cross-sections, MIR Si strip waveguides are multi-mode at NIR wavelengths, hindering device implementation. Here we present, for the first time, an integrated Bragg grating waveguide filter for hybrid near- and mid-infrared operation. Specifically, the filter is implemented in a single-etch suspended silicon corrugated waveguide with an effectively single-mode operation in NIR region for a waveguide cross-section as large as $0.5\,\mu\mathrm{m}\,\times\, 1.1\,\mu\mathrm{m}$. At the same time, the waveguide supports single-mode propagation in MIR region. We demonstrate a long-period waveguide Bragg grating yielding a sharp third-order Bragg resonance for the fundamental waveguide mode and radiating the higher order modes. We experimentally demonstrate a notch filter with a $4\,\mathrm{nm}$ bandwidth and $40\,\mathrm{dB}$ extinction ratio with a temperature-dependent Bragg wavelength shift of $70\,\mathrm{pm/K}$.
	\end{abstract}
	\maketitle
	
	The mid-infrared (MIR) wavelength range contains the vibrational and rotational resonances of most of the molecules and MIR spectroscopy is widely used to identify and analyze chemical and biological substances. Advances in high-performance quantum cascade lasers \cite{QCL_review} have opened new opportunities for the development of compact MIR spectrometers operating at room temperature. Albeit good-quality Peltier-cooled MCT detectors are commercially available \cite{Peltier_MCT}, high-performance free-space MIR absorption spectrometers often rely on cryogenic-cooled MIR photodetectors, which are difficult to integrate. An alternative approach exploits the photothermal effect to indirectly detect MIR absorption \cite{Generic_PTE}. Through interaction with the target molecules, MIR light is absorbed and converted into heat, changing material properties, e.g. refractive index. By monitoring these changes by using NIR wavelengths, it is possible to measure MIR absorption, obviating the need for MIR photodetectors. Photo-acoustic spectrometers based on the photothermal effect have experimentally shown a superior sensitivity compared to direct transmission spectroscopy \cite{PTE_Performance_A,PTE_Performance_B,PTE_Performance_C}. Integration of photothermal sensors has been proposed, relying on MIR pump and probe light \cite{PTE_JJ_1}. However, this approach still requires an MIR photodetector. First demonstration of photothermal effect in Si ring resonators has recently been reported where NIR shift is induced by a surface-illuminating MIR pump \cite{PTE_Ghent}. A promising alternative integrated photothermal-based approach has been proposed where both MIR and NIR light resonate in the same integrated photonic microcavity \cite{PTE_JJ_2}. The heat generated by the absorption of pump MIR light rises the temperature of the cavity, changing its effective index, and shifting the resonance wavelength of the probe NIR light. By monitoring this NIR resonance shift, it is possible to estimate the MIR absorption, without the need for high-performance MIR photodetectors. Temperature rise in the cavity is limited by thermal leakage, which is proportional to the material thermal conductivity. On the other hand, the resonance shift is proportional to the variation of cavity effective index with temperature ($\mathrm{d} n_{eff}/\mathrm{d} T$).

	\begin{figure}[htbp]
		\centering
		\centerline{\includegraphics[width=5cm]{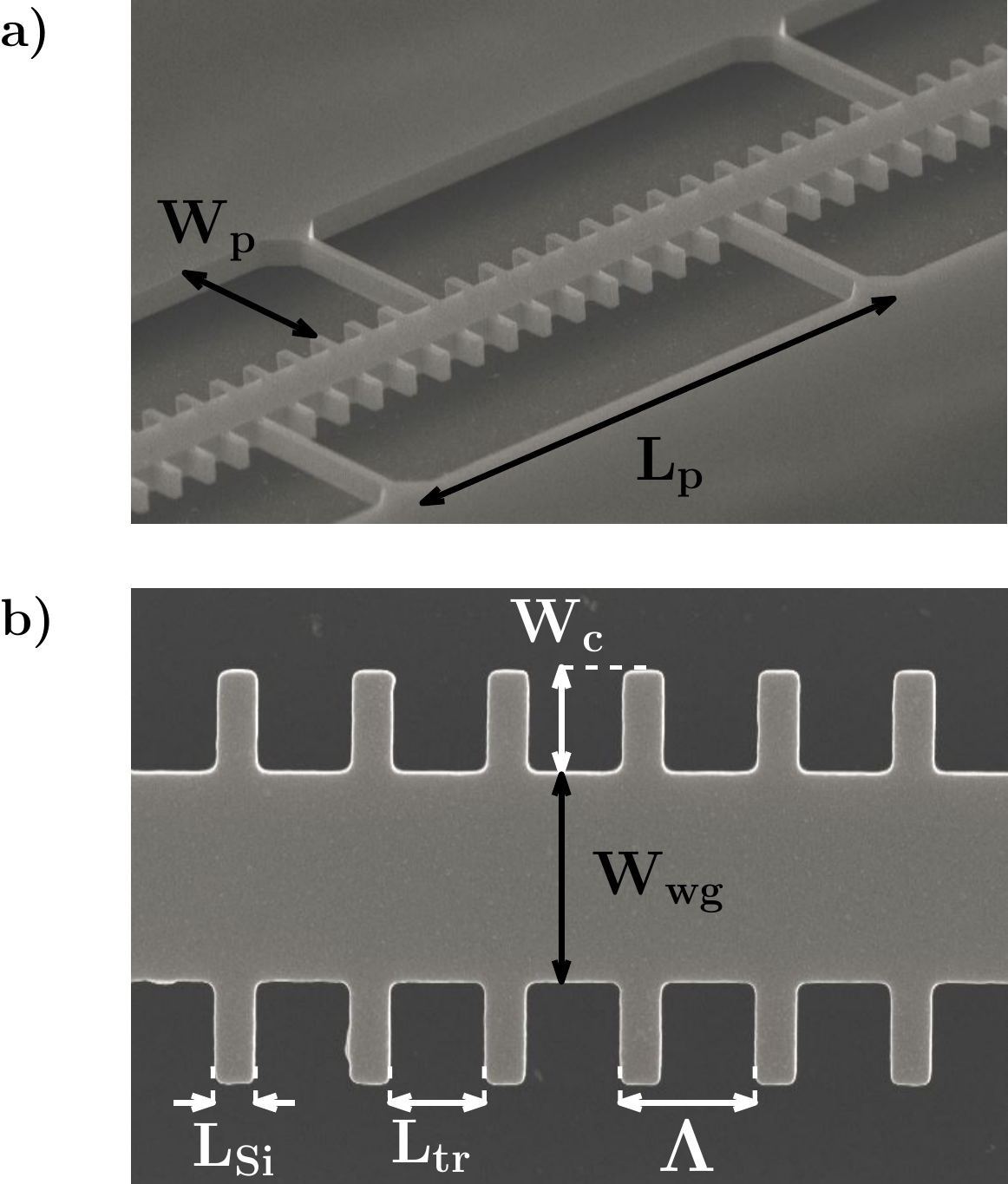}}
		\caption{Scanning electron microscope (SEM) image of the fabricated suspended silicon $3^{\mathrm{rd}}$-order Bragg filter. (a) Side view showing anchoring pillars and (b) detail of the filter geometry.}
		\label{fig:FabrSEMs}
	\end{figure}
	
	Group-IV materials, particularly silicon and germanium, are advantageously used in mature microelectronic facilities for cost-effective production of photonic devices at large volumes. While the germanium-on-silicon platform has shown very low propagation loss at MIR wavelengths \cite{GOS_LowLoss}, its NIR transparency is limited by Ge absorption below $1.6\,\mu\mathrm{m}$. On the other hand, the Si transparency range (from $1.1\,\mu\mathrm{m}$ to $8\,\mu\mathrm{m}$) covers both the NIR wavelengths and an important part of the MIR region. The utilization of silicon-on-insulator (SOI) platform in the MIR region is nevertheless restricted by a large absorption of buried $\mathrm{SiO_2}$ layer for wavelengths above $4\,\mu\mathrm{m}$. The use of substrates with wider transparency ranges, such as silicon nitride \cite{SON_A} or sapphire \cite{SOS_B}, has been proposed to overcome this limitation. Another promising approach is the removal of the buried oxide yielding membrane structures \cite{Susp_Si,Susp_Si_Ours}. The membrane Si waveguides can cover the full transparency window of silicon, yet benefiting from the high quality materials and mature silicon fabrication processes. However, silicon exhibits a thermal conductivity of 149 W/mK, which is substantially larger than that of chalcogenide glass waveguides (0.22 W/mK) proposed in \cite{PTE_JJ_2}, resulting in potentially lower photothermal sensitivity. Still, the access to mature fabrication processes and high performance integrated Ge NIR photodetectors \cite{Ge_PD}, make pump-probe Si photothermal spectrometers very interesting for applications where the cost is more important than the maximum sensitivity \cite{FoodControl}.   
	
	The performance of photothermal absorption spectrometers based on MIR pump / NIR probe configuration relates to their ability to resolve small variations in the NIR transmission spectrum. Suspended Si waveguides with single-mode behavior for both NIR and MIR wavelengths are needed for such sensors. However, this is a challenging requirement due to the high index contrast between the Si core and the air cladding. Specifically, conventional strip waveguides with cross-section large enough to guide a MIR mode (e.g. within the atmospheric transparency window in the wavelength range $3-5\,\mu\mathrm{m}$) are multi-mode in the NIR range. Alternatively, rib waveguides can exhibit an ultra-wideband single-mode behavior \cite{Soref_Rib} which could cover both the NIR and MIR. Nevertheless, suspended rib waveguides require two etching steps that complicate device fabrication \cite{Susp_Si}. In addition, the modal field extending in the lateral slabs of the rib compromises the MIR pump light interaction with the target molecules, thus reducing device sensitivity. 
	
	In this Letter, we propose a new corrugated waveguide geometry (see Fig. \ref{fig:FabrSEMs}) that suppresses diffraction effects for a grating period substantially larger than conventional sub-wavelength design. We exploit this concept to afford single-mode propagation at NIR wavelengths (near $1.55\,\mu\mathrm{m}$) for a waveguide having a cross-section as large as $550\,\mathrm{nm}$ in thickness by $1.1\,\mu\mathrm{m}$ in width. The purpose of the grating is two-fold. First, the grating radiates all the higher-order modes to ensure single-mode behavior. Second, it creates a sharp $3^{\mathrm{rd}}$-order Bragg reflection for the fundamental mode, which allows temperature monitoring. Advantageously, this corrugated waveguide also provides single-mode operation at MIR wavelengths near $3.8\,\mu\mathrm{m}$, opening a new route for the implementation of high performance pump-probe photothermal absorption spectrometers.          	    
	
	We choose a Si thickness of $500\,\mathrm{nm}$ for compatibility with previously developed MIR suspended waveguide structures \cite{Susp_Si_Ours,Susp_Si_Ours_2}. The periodic waveguide (see Fig. \ref{fig:FabrSEMs}(b)) has a width of $W_{wg}$ and a lateral corrugation depth of $W_c$. The grating teeth and trench lengths are $L_{Si}$ and $L_{tr}$, respectively. The period of the structure is $\Lambda=L_{Si}+L_{tr}$ and the duty cycle is defined as $DC=L_{Si}/\Lambda$.
	
	\begin{figure}[htbp]
		\centering
		\centerline{\includegraphics[width=5.75cm]{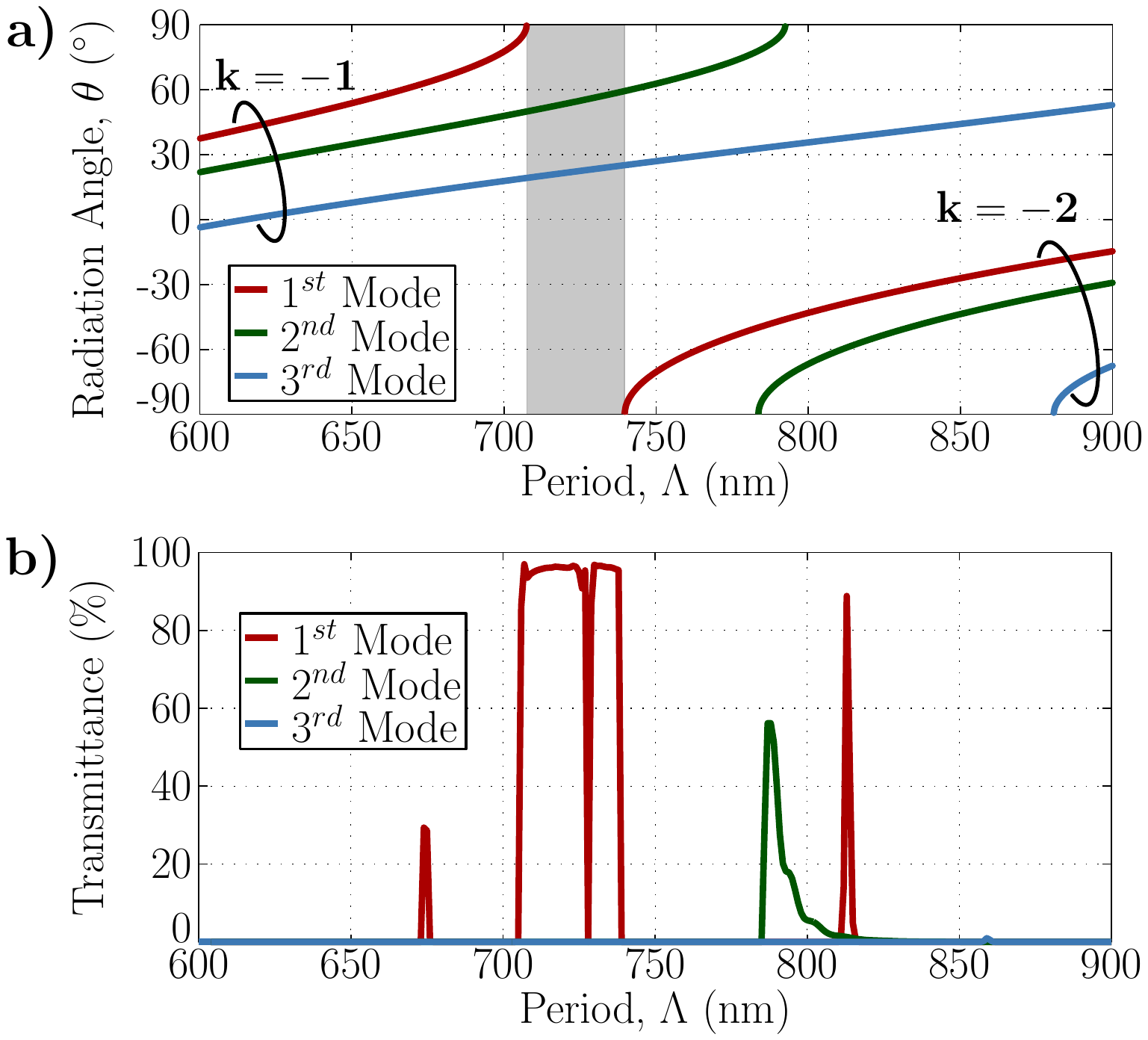}}
		\caption{(a) Radiation angle (Eq. (\ref{eq:RadCondition})), as a function of the corrugation period for the $k=-1$ and $k=-2$ diffraction orders of the first three waveguide modes for $1.55\,\mu\mathrm{m}$ wavelength. (b) Calculated transmittance for $1.55\,\mu\mathrm{m}$ wavelength as a function of the period for the first three modes of a 1-cm-long waveguide with $W_c=500\,\mathrm{nm}$ and $DC=0.3$.}
		\label{fig:PitchStudy}
	\end{figure}
	
	Radiation in a periodic structure is allowed for diffraction orders meeting the phase-matching condition \cite{Tamir}:     
	\begin{equation}
	\left| \sin\left(\theta_{k}\right) \right|= \left|\frac{n_{B}}{n_{c}} + \frac{k \lambda}{n_{c}\Lambda}\right|\leqslant 1,
	\label{eq:RadCondition}
	\end{equation}
	where $k$ is the diffraction order, $\lambda$ is the wavelength, $n_{B}$ is the effective index of the Bloch-Floquet mode in the periodic waveguide, $\theta_{k}$ is the radiation angle and $n_{c}$ is the refractive index of the waveguide cladding. For example, fiber-chip surface grating couplers typically operate by making the first ($k=-1$) \cite{GC_BottomMirror,SWG_Grat_A,SWG_Grat_B} or second ($k=-2$) \cite{InvTaperExp} diffraction order meet the radiation condition. Conversely, sub-wavelength grating waveguides suppress diffraction by choosing a period $\Lambda<\lambda/(2n_B)$ \cite{FirstSwg,SwgReview}. Interestingly, from Eq. (\ref{eq:RadCondition}) it follows that for $n_B>3n_c$, a diffraction-less region exists for: $\lambda/(n_B-n_c)<\Lambda<2\lambda/(n_B+n_c)$, since no diffraction orders are allowed in that range. This is the fundamental principle which we exploit in this work. This operation regime offers some important practical benefits. The diffraction effects are suppressed for a grating period approximately 3 times larger than the conventional sub-wavelength grating design. The longer non-radiative periods relax the fabrication constraints and widen the duty cycles that can be fabricated for a given minimum feature size. Furthermore, our design strategy allows to engineer the width of a multi-mode waveguide to make only the fundamental mode (with the highest effective index) meet the condition $n_B>3n_c$, for which diffraction is suppressed. Hence, high-order modes will be radiated out by the periodic structure, yielding an effectively single-mode behavior.

	Figure \ref{fig:PitchStudy}(a) shows the radiation angle $\theta_k$ calculated from Eq. (\ref{eq:RadCondition}) as a function of the corrugation period ($\Lambda$) for different waveguide modes and diffraction orders. Here, we considered a waveguide width of $W_{wg}=1.1\,\mu\mathrm{m}$, a wavelength of $1.55\,\mu\mathrm{m}$ and transverse-electric (TE) polarized light. We approximated the indices of the Bloch-Floquet modes in the periodic structure by effective indices of the equivalent unperturbed waveguide, yielding $n_{B_1}=3.2$, $n_{B_2}=2.9$ and $n_{B_3}=2.5$ for the first, the second and the third modes, respectively. It is observed that the fundamental mode, with $n_{B_1}>3n_c=3$, propagates diffraction-less in the shadowed region of Fig. \ref{fig:PitchStudy}(a) where no radiating diffraction orders exist. We also studied light propagation in this corrugated waveguide at $1.55\,\mu\mathrm{m}$ wavelength using a 2-D Fourier expansion simulation tool with Bloch-Floquet formalism \cite{FEXEN}. The waveguide length and corrugation depth are $1\,\mathrm{cm}$ and $W_c=500\,\mathrm{nm}$, respectively, and the duty cycle is $DC=0.3$. Figure \ref{fig:PitchStudy}(b) shows calculated transmittance for the first three waveguide modes. The fundamental mode exhibits high transmittance for grating periods within the diffraction-less region, while higher order modes are radiated out. Therefore, effectively single-mode behavior is achieved.

	\begin{figure}[htbp]
		\centering
		\centerline{\includegraphics[width=6cm]{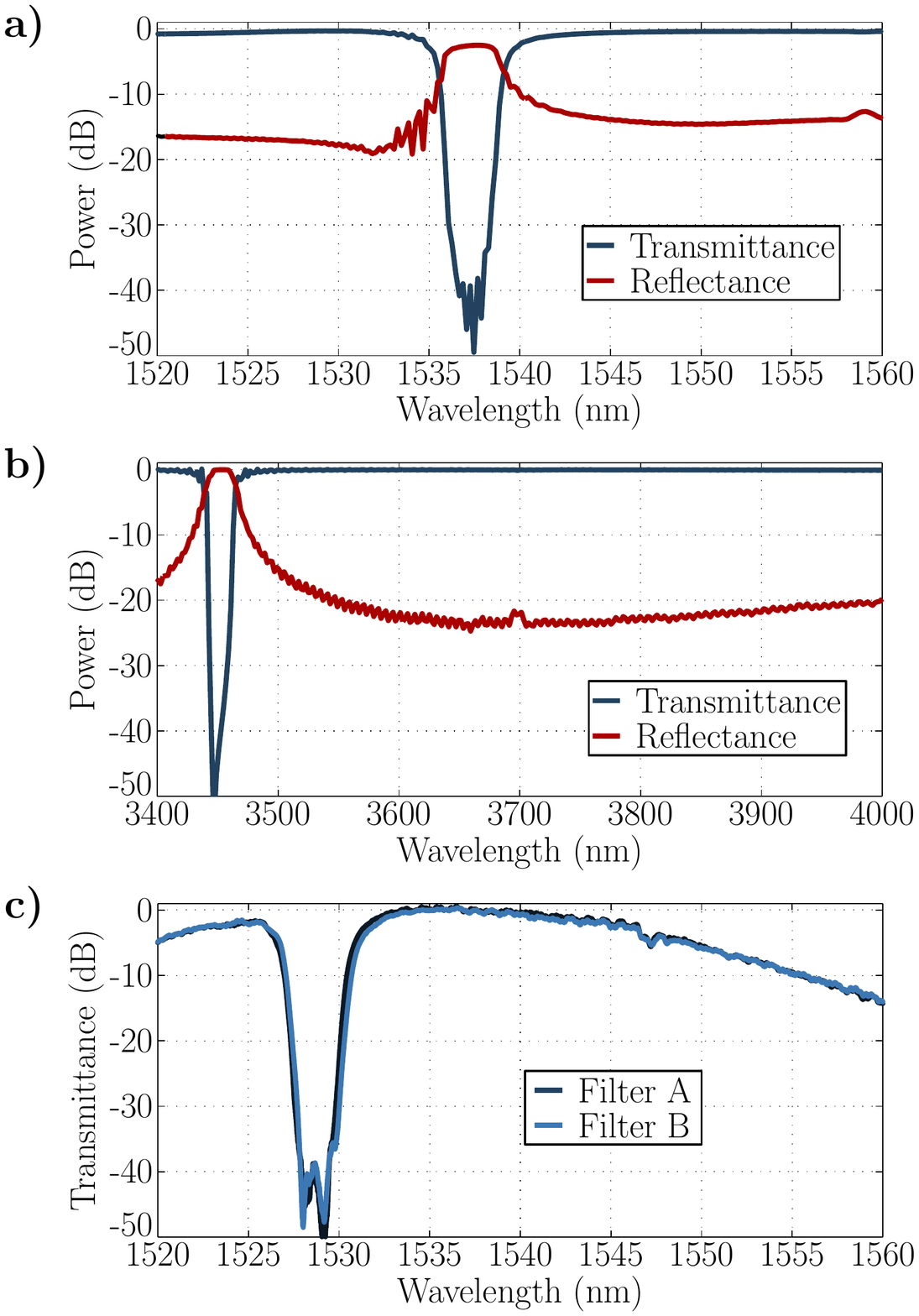}}
		\caption{Filter spectra calculated with 3D-FDTD for (a) NIR and (b) MIR bands. (c) Measured filter transmittance in the NIR region.  $500$-$\mu\mathrm{m}$-long corrugated waveguide with $\Lambda=725\,\mathrm{nm}$, $DC=0.3$, $W_c=500\,\mathrm{nm}$ and $W_{wg}=1.1\,\mu\mathrm{m}$.}
		\label{fig:TransmAnalysis}
	\end{figure}

	The low-transmittance notch within the diffraction-less region in Fig. \ref{fig:PitchStudy}(b) corresponds to the third-order Bragg resonance for $\Lambda=(3\lambda)/(2n_{B_1})$. By monitoring the thermally induced shift of the third-order Bragg NIR resonance one can monitor temperature variations due to the photothermal effect. Bandwidth of the Bragg resonance is proportional to the grating corrugation depth \cite{Formulas}. Thus, shallow corrugations are preferred for narrow-band resonant features, yielding an increased spectral sensitivity. Figure \ref{fig:TransmAnalysis}(a) shows the transmitted and the reflected power as a function of the wavelength, calculated with the three-dimensional (3-D) Finite Difference Time Domain (FDTD) simulator from Lumerical Solutions. The filter length is $500\,\mu\mathrm{m}$, the period is $\Lambda=725\,\mathrm{nm}$, the duty cycle is $DC=0.3$, the waveguide width is $W_{wg}=1.1\,\mu\mathrm{m}$ and grating corrugation depth is $W_c=500\,\mathrm{nm}$. We used the fundamental TE mode of the unperturbed waveguide as an excitation filed and included $10$-$\mu\mathrm{m}$-long adiabatic transitions between the strip and the corrugated waveguides. Calculations predict a $40\,\mathrm{dB}$ reflection notch with a narrow $3\,\mathrm{dB}$ bandwidth of $3.15\,\mathrm{nm}$ for a corrugation depth of $W_c=500\,\mathrm{nm}$, corresponding to a Bragg coupling coefficient of $k=95\,\mathrm{mm^{-1}}$. For comparison, conventional Bragg filters in Si-wires with similar bandwidth require corrugation depths of about $30\,\mathrm{nm}$. In addition, the Bloch-Floquet mode of our suspended Si Bragg filter can leverage the high Si thermo-optic coefficient ($2 \times 10^{-4}\mathrm{/K}$) to yield an effective index variation with temperature of $1.9\times 10^{-4}\,\mathrm{/K}$. For comparison, TE Si-wire waveguides used in high-performance temperature sensors operating near $1.55\,\mu\mathrm{m}$ wavelength exhibit similar effective index variations with temperature \cite{RR_T_Sensor}.              
	
	Suspended Si waveguides with sub-wavelength cladding have been demonstrated with propagation loss as low as $0.82\,\mathrm{dB/cm}$ at $3.8\,\mu\mathrm{m}$ wavelength \cite{Susp_Si_Ours_2}. These waveguides yield single-mode MIR propagation for waveguide cross-sections similar to those proposed here. Interestingly, the $725\,\mathrm{nm}$ periodicity, that provides diffraction-less operation in NIR region, yields a first-order Bragg wavelength of $3.45\,\mu\mathrm{m}$ (see Fig. \ref{fig:TransmAnalysis}(b)). Hence, it operates in the sub-wavelength regime for wavelengths near $3.8\,\mu\mathrm{m}$. Therefore, our suspended Si periodic structure can serve as a generic low-loss waveguide in MIR range.
	
	We fabricated a set of suspended Si filters with different parameters to experimentally validate their NIR effectively single-mode behavior. We used a SOI platform with $500$-$\mathrm{nm}$-thick Si layer and $2$-$\mu \mathrm{m}$-thick buried oxide layer. As shown in the scanning electron microscope (SEM) image in Fig. \ref{fig:FabrSEMs}(a), we used a series of equidistantly spaced pillars to anchor the corrugated waveguide to the lateral Si slab. A pillar width of $W_{p}=3.5\,\mu\mathrm{m}$ was used to ensure a sufficient optical isolation between the waveguide mode and the lateral silicon slab regions. The pillar separation of $L_p=10.15\,\mu\mathrm{m}$ ($14\times\Lambda$) ensures sufficient mechanical stability, while minimizing perturbation of the filter response. Input and output sub-wavelength grating couplers \cite{SWG_Grat_A,SWG_Grat_B} were optimized to minimize Bragg reflections that could distort measured spectral response \cite{WeiWeiRings}. Grating couplers were connected to the filters by a short multi-mode strip section and an adiabatic taper. We used electron beam lithography (Nanobeam NB-4 system, 80 kV) and dry etching with an inductively coupled plasma etcher ($\mathrm{SF_6}$ gas) for structure definition. We removed the buried oxide layer to form the suspended structure by an 1 hour exposure to hydrofluoric acid vapor.
	
	Figure \ref{fig:TransmAnalysis}(c) shows measured spectrum for two nominally identical filters with waveguide width $W_{wg}=1.1\,\mu\mathrm{m}$, corrugation depth $W_c=500\,\mathrm{nm}$, grating period $\Lambda=725\,\mathrm{nm}$, duty cycle $DC=0.3$ and filter length $500\,\mu\mathrm{m}$. The filter response shows no signature of multi-mode beating, with a clear notch at $1529\,\mathrm{nm}$ wavelength, confirming the predicted effectively single-mode behavior. The experimental bandwidth ($\sim 4\,\mathrm{nm}$) and depth ($\sim 40\,\mathrm{dB}$) of the Bragg resonance are in good agreement with the simulation results. The blue-shift of the notch can be attributed to slight over-etching of the fabricated devices. The envelope of the measured spectrum is determined by the wavelength-dependent coupling efficiency of the input and the output grating couplers.
	
\begin{figure}[htbp]
		\centering
		\centerline{\includegraphics[width=6cm]{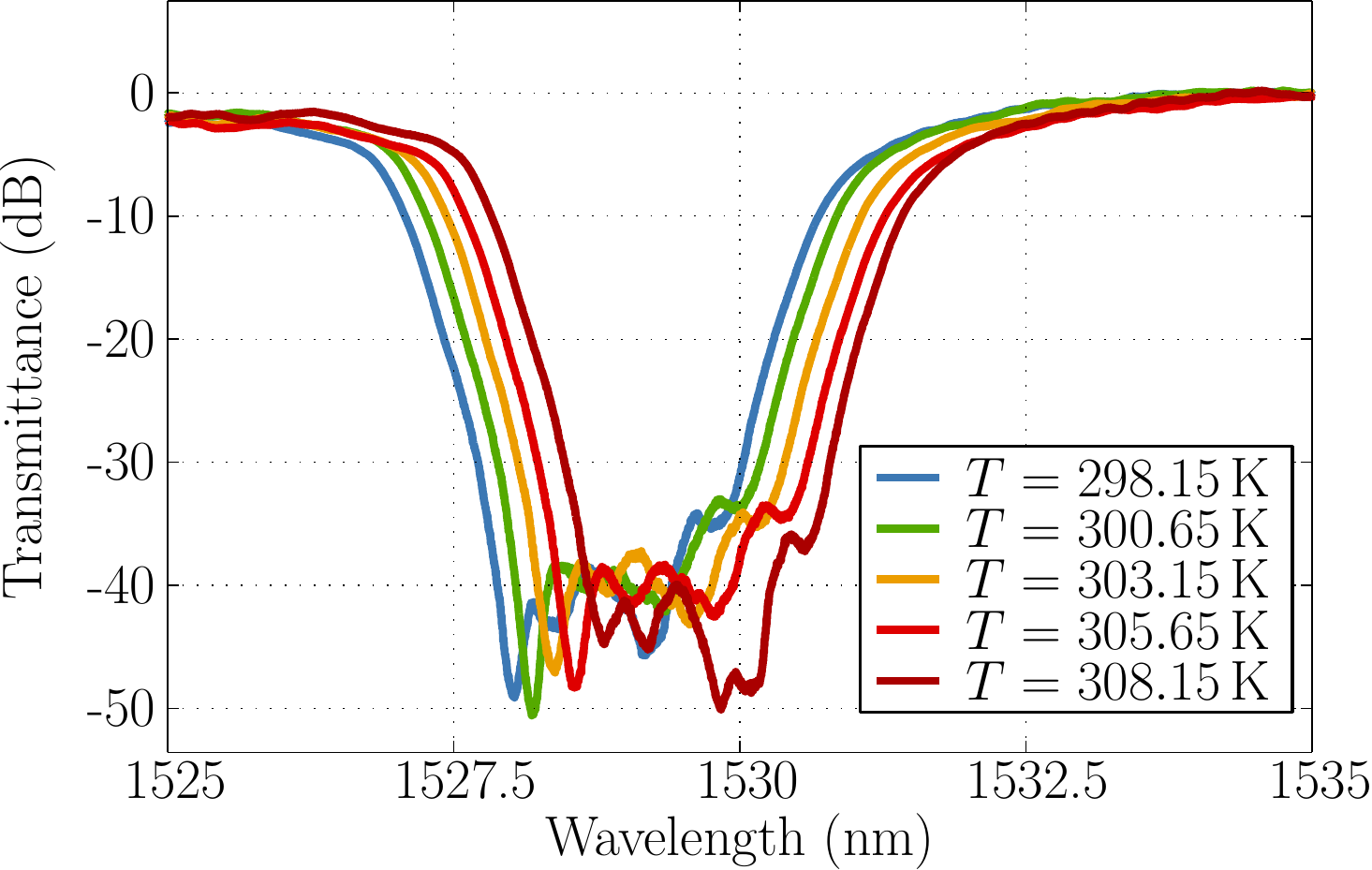}}
		\caption{Measured transmission spectra of the fabricated Bragg filter at different temperatures. $500$-$\mu\mathrm{m}$-long corrugated waveguide with $\Lambda=725\,\mathrm{nm}$, $DC=0.3$, $W_c=500\,\mathrm{nm}$ and $W_{wg}=1.1\,\mu\mathrm{m}$.}
		\label{fig:TempAnalysis}
	\end{figure}

	Finally, we experimentally evaluate the variation of the filter effective index with the temperature ($\mathrm{d} n_{eff}/\mathrm{d} T$) and the temperature dependent wavelength shift, which are key parameters determining the potential sensitivity of photothermal absorption spectrometers. To perform optical characterization, we used a Peltier system to control filter temperature while monitoring the position of the Bragg resonance. Figure \ref{fig:TempAnalysis} shows the measured spectrum of our $500$-$\mu\mathrm{m}$-long filter for different temperatures. The Bragg resonance wavelength shifts nearly linearly with a coefficient of $\sim70\,\mathrm{pm/K}$, similar to that of state-of-the-art temperature sensors implemented with Si-wires \cite{RR_T_Sensor}. The temperature-dependent wavelength variation is used to estimate the corresponding effective index variation with temperature. This yields $\mathrm{d} n_{eff}/\mathrm{d} T\,\sim 1.5\times10^{-4}\mathrm{/K}$, smaller than the theoretical value $1.9\times10^{-4}\mathrm{/K}$. This difference may be attributed to a high thermal isolation between the suspended Si filter and the Si substrate. This may result in a lower temperature in the waveguide compared to the Si substrate (controlled by the Peltier system), thus reducing temperature-dependent shift of the Bragg resonance.

	In conclusion, we proposed and experimentally demonstrated, for the first time, a suspended Si waveguide with a corrugated structure to suppress propagation of higher-order modes. This new waveguide geometry yields effectively single-mode operation at NIR wavelengths, while at the same time provides a cross-section large enough to guide MIR wavelengths. In addition, the periodic corrugation creates a narrow-band third-order Bragg resonance. This property can be advantageously used to monitor temperature changes due to the photothermal effect. Based on this concept, we designed and experimentally demonstrated notch filters in single-etch suspended Si waveguides with cross-sections as large as $0.5\,\mu\mathrm{m}\times 1.1\,\mu\mathrm{m}$. The measured spectral response of the filter shows no signature of multi-mode propagation, with a $40\,\mathrm{dB}$ Bragg resonance and remarkably narrow bandwidth of $\sim4\,\mathrm{nm}$ for a $500\,\mathrm{nm}$ deep corrugation. At the same time, we experimentally demonstrated a temperature-dependent Bragg resonance shift of $\sim70\,\mathrm{pm/K}$ comparable to state-of-the-art Si-wire temperature sensors \cite{RR_T_Sensor}. These results open a new path towards the realization of integrated silicon-based pump-probe photothermal absorption spectrometers that would obviate the need for MIR photodetectors.


\begin{thebibliography}{10}
		\newcommand{\enquote}[1]{``#1''}
		
		\bibitem{QCL_review}
		M. Razeghi, Q. Y. Lu, N. Bandyopadhyay, W. Zhou, D. Heydari, Y. Bai, and S. Slivken, ``Quantum cascade lasers: from tool to product,'' Opt. Express \textbf{23}, 8462-8475 (2015).
		
		\bibitem{Peltier_MCT}
		http://www.vigo.com.pl/products/infrared-detectors
		
		\bibitem{Generic_PTE}
		L.A. Skvortsov, ``Laser photothermal spectroscopy of light-induced absorption,'' Quantum Electronics \textbf{43}, 1-13 (2013).
		
		\bibitem{PTE_Performance_A}
		V. Spagnolo, P. Patimisco, S. Borri, G. Scamarcio, B. E. Bernacki, and J. Kriesel, ``Part-per-trillion level $\mathrm{SF_6}$ detection using a
		quartz enhanced photo-acoustic spectroscopy-based sensor with single-mode fiber-coupled quantum cascade laser excitation,'' Opt. Lett. \textbf{37}, 4461-4463 (2012).
		
		
		\bibitem{PTE_Performance_B}
		Y. Zhao, K. Liu, J. McClelland, and M. Lu, ``Enhanced photoacoustic detection using photonic crystal substrate,'' Appl. Phys. Lett. \textbf{114}, 161110-1-161110-5 (2014).
		
		\bibitem{PTE_Performance_C}
		J. Peltola, T. Hieta, AND M. Vainio, ``Parts-per-trillion-level detection of nitrogen dioxide by cantilever-enhanced photo-acoustic spectroscopy,'' Opt. Lett. \textbf{40}, 2933-2936 (2015).
		
		
		
		\bibitem{PTE_JJ_1}
		J. Hu, ``Ultra-sensitive chemical vapor detection using micro-cavity photothermal spectroscopy,'' Opt. Express \textbf{18}, 22174-22186 (2010).
		
		\bibitem{PTE_Ghent}
		A. Vasiliev, A. Malik, M. Muneeb, R. Baets and G. Roelkens, ``Photothermal mid-Infrared spectroscopy using Fano resonances in silicon microring resonators,'' in Conference on Laser and Optoelectronics Europe (CLEO) (OSA, 2016) paper SF2H.5.
		
		\bibitem{PTE_JJ_2}
		H. Lin, Z. Yi, and J. Hu, ``Double resonance 1-D photonic crystal cavities for single-molecule mid-infrared photothermal spectroscopy: theory and design,'' Opt. Lett. \textbf{37}, 1304-1306 (2012).
		
		\bibitem{GOS_LowLoss}
		M. Nedeljkovic, J. Soler Penad\'{e}s, C. J. Mitchell, A. Z. Khokhar, S. Stankovi\'{c}, T. Dominguez Bucio, C. G. Littlejohns, F. Y. Gardes, and G. Z. Mashanovich, ``Surface-grating-coupled low-loss Ge-on-Si rib waveguides and multimode interferometers,'' IEEE Photon. Technol. Lett. \textbf{27}, 1040-1043 (2015).
		
		\bibitem{SON_A}
		J. Mu, R. Soref, L. C. Kimerling, and J. Michel, ``Silicon-on-nitride structures for mid-infrared gap-plasmon waveguiding,'' Appl. Phys. Lett. \textbf{104}, 031115 (2014).
		
		\bibitem{SOS_B}
		N. Singh, D. D. Hudson, Y. Yu, Ch. Grillet, S. D. Jackson, A. Casas-Bedoya, A. Read, P. Atanackovic, S. G. Duvall, S. Palomba, B. Luther-Davies, S. Madden, D. J. Moss, and B. J. Eggleton, ``Mid-infrared supercontinuum generation from 2 to 6 $\mu \mathrm{m}$ in a silicon nanowire,'' Optica  \textbf{2}, 797-802 (2015).
		
		\bibitem{Susp_Si}
		X. Wang, Z. Cheng, K. Xu, H. K. Tsang, and J.-B. Xu, ``High-responsivity graphene/silicon-heterostructure waveguide photodetectors,'' Nature Photon.  \textbf{7}, 888-891 (2013).
		
		\bibitem{Susp_Si_Ours}
		J. Soler Penad\'{e}s, C. Alonso-Ramos, A. Z. Khokhar, M. Nedeljkovic, L. A. Boodhoo, A. Ortega-Mo\~{n}ux, I. Molina-Fern\'{a}ndez, P. Cheben, and G. Z. Mashanovich, ``Suspended SOI waveguide with sub-wavelength grating cladding for mid-infrared,'' Opt. Lett.  \textbf{39}, 5661-5664 (2014).
		
		\bibitem{Susp_Si_Ours_2}
		J. Soler Penad\'{e}s, A. Ortega-Mo\~{n}ux, M. Nedeljkovic, J. G. Wang{\"u}emert-P\'{e}rez, R. Halir, A. Z. Khokhar, C. Alonso-Ramos, Z. Qu, I. Molina-Fern\'{a}ndez, P. Cheben, and G. Z. Mashanovich, ``Suspended silicon mid-infrared waveguide devices with subwavelength grating metamaterial cladding,'' Opt. Express  \emph{Accepted}.
		
		
		\bibitem{Ge_PD}
		L. Virot, P. Crozat, J. F\'{e}d\'{e}li, J. Hartmann, D. Marris-Morini, E. Cassan, F. Boeuf, and L. Vivien, ``Germanium avalanche receiver for low power interconnects,'' Nat. Commun.  \textbf{5}, 1-6 (2014).
		
		
		\bibitem{FoodControl}
		S. Neethirajan, D.S. Jayas, S. Sadistap, ``Carbon dioxide ($\mathrm{CO_2}$) sensors for the agri-food industry - a review,'' Food Bioprocess Technol.  \textbf{2}, 115-121 (2009).
		
		\bibitem{Soref_Rib}
		R. Soref, J. Schmidtchen, and K. Petermann, ``Large single-mode rib waveguides in GeSi-Si and Si-on-SiO$_2$,'' IEEE J. Quantum Electron.  \textbf{27}, 1971-1974 (1991).
		
		\bibitem{Tamir}
		T. Tamir, and S. Peng, ``Analysis and design of grating couplers,'' Appl. Phys. A-Mater. \textbf{14}, 235-254 (1977).
		
		
		\bibitem{GC_BottomMirror}
		W. S. Zaoui, A. Kunze, W. Vogel, M. Berroth, J. Butschke, F. Letzkus, and J. Burghartz, ``Bridging the gap between optical fibers and silicon photonic integrated circuits,''  Opt. Express \textbf{22}, 1277-1286 (2014).
		
		
		\bibitem{SWG_Grat_A}
		R. Halir, P. Cheben, S. Janz, D.-X. Xu, I. Molina-Fern\'{a}ndez, and J. G. Wang{\"u}emert-P\'{e}rez, ``Waveguide grating coupler with subwavelength microstructures,'' Opt. Lett. \textbf{34}, 1408-1410 (2009).
		
		
		\bibitem{SWG_Grat_B}
		D. Benedikovic, C. Alonso-Ramos, P. Cheben, J. H. Schmid, S. Wang, R. Halir, A. Ortega-Mo\~{n}ux, D.-X. Xu, L. Vivien, J. Lapointe, S. Janz, and M. Dado, ``Single-etch subwavelength engineered fiber-chip grating couplers for $1.3\,\mu\mathrm{m}$ datacom wavelength band,'' Opt. Express \textbf{24}, 12893-12904 (2016).
		
		
		
		\bibitem{InvTaperExp}
		C. Alonso-Ramos, A. Ortega-Mo\~{n}ux, L. Zavargo-Peche, R. Halir, J. de Oliva-Rubio, I. Molina-Fern\'{a}ndez, P. Cheben, D.-X. Xu, S. Janz, N. Kim, and B. Lamontagne, ``Single-etch grating coupler for micrometric silicon rib waveguides,'' Opt. Lett. \textbf{36}, 2647-2649 (2011).
		
		\bibitem{FirstSwg}
		P. Cheben, P. J. Bock, J. H. Schmid, J. Lapointe, S. Janz, D.-X. Xu, A. Densmore, A. Del\^{a}ge, B. Lamontagne, and T. J. Hall, ``Refractive index engineering with subwavelength gratings for efficient microphotonic couplers and planar waveguide multiplexers,'' Opt. Lett. \textbf{35}(15), 2526-2528 (2010).
		
		\bibitem{SwgReview}
		R. Halir, P. J. Bock, P. Cheben, A. Ortega-Mo\~{n}ux, C. Alonso-Ramos, J. H. Schmid, J. Lapointe, D.-X. Xu, J. G. Wang{\"u}emert-P{\'e}rez, I. Molina-Fern\'{a}ndez and S. Janz ``Waveguide sub-wavelength structures: a review of principles and applications,'' Laser Photonics Rev. \textbf{35}, 25-49 (2015).
		
		\bibitem{FEXEN}
		L.  Zavargo-Peche, A. Ortega-Mo\~{n}ux,  J.  G.  Wang{\"u}emert-P\'{e}rez, and I. Molina-Fern\'{a}ndez, ``Fourier  based  combined  techniques  to  design  novel sub-wavelength  optical  integrated  devices,''  Prog.  Electromagn.  Res. \textbf{123}, 447-465, (2012). 
		
		
		\bibitem{Formulas}
		D. C. Flanders, H. Kogelnik, R. V. Schmidt, and C. V. Shank, ``Grating filters for thin-film optical waveguides,'' Appl. Phys. Lett. \textbf{24}(4), 194-196 (1974).
		
		\bibitem{RR_T_Sensor}
		H. Xu, M. Hafezi, J. Fan, J. M. Taylor, G. F. Strouse, and Z. Ahmed, ``Ultra-sensitive chip-based photonic temperature sensor using ring resonator structures,'' Opt. Express \textbf{22}, 3098-3104 (2014).
		
		
		
		\bibitem{WeiWeiRings}
		W. Zhang, S. Serna, X. Le Roux, C. Alonso-Ramos, L. Vivien, and E. Cassan, ``Analysis of silicon-on-insulator slot waveguide ring resonators targeting high Q-factors,'' Opt. Lett. \textbf{40}, 5566-5569 (2015).
		
		
		
	\end{thebibliography}
\end{document}